# Sparse LMS via Online Linearized Bregman Iteration

Tao Hu, and Dmitri B. Chklovskii
Howard Hughes Medical Institute, Janelia Farm Research Campus
{hut, mitya}@janelia.hhmi.org

***Abstract*—We propose a version of least-mean-square (LMS) algorithm for sparse system identification. Our algorithm called online linearized Bregman iteration (OLBI) is derived from minimizing the cumulative prediction error squared along with an $l_1$-$l_2$ norm regularizer. By systematically treating the non-differentiable regularizer we arrive at a simple two-step iteration. We demonstrate that OLBI is bias free and compare its operation with existing sparse LMS algorithms by rederiving them in the online convex optimization framework. We perform convergence analysis of OLBI for white input signals and derive theoretical expressions for both the steady state and instantaneous mean square deviations (MSD). We demonstrate numerically that OLBI improves the performance of LMS type algorithms for signals generated from sparse tap weights.**

## I. Introduction

MANY signal processing tasks, such as signal prediction, noise cancellation or system identification, reduce to predicting a time varying signal, $f_k$ [1]. In many cases such prediction can be computed as a linear combination of the input signal vector, $\boldsymbol{x}_k$, with the weight vector, $\boldsymbol{w}_k$. The goal is to minimize the cumulative empirical loss defined as the square of the prediction error,

$$\sum_{k=1}^{t} \ell_k(f_k, \boldsymbol{w}_k) = \sum_{k=1}^{t} \frac{1}{2}(f_k - \boldsymbol{w}_k^T \boldsymbol{x}_k)^2, \quad (1)$$

by continuously adjusting the weights, $\boldsymbol{w}_k$, based on the previous prediction error. This adaptive filtering framework, Fig. 1, can be formalized as follows:

Initialize the weight vector, $\boldsymbol{w}_1$. At each time step $k = 1,2, \dots, t$

- Get input signal vector, $\boldsymbol{x}_k = [x_{k,0}, x_{k,1}, \dots, x_{k,n-1}]^T$.
- Compute a filter output, $\boldsymbol{w}_k^T \boldsymbol{x}_k$, for the desired output, $f_k$, using the weight vector, $\boldsymbol{w}_k = [w_{k,0}, w_{k,1}, \dots, w_{k,n-1}]^T$.
- Observe desired output signal, $f_k$.
- Adjust weight vector $\boldsymbol{w}_k \Rightarrow \boldsymbol{w}_{k+1}$, using prediction error, $f_k - \boldsymbol{w}_k^T \boldsymbol{x}_k$.

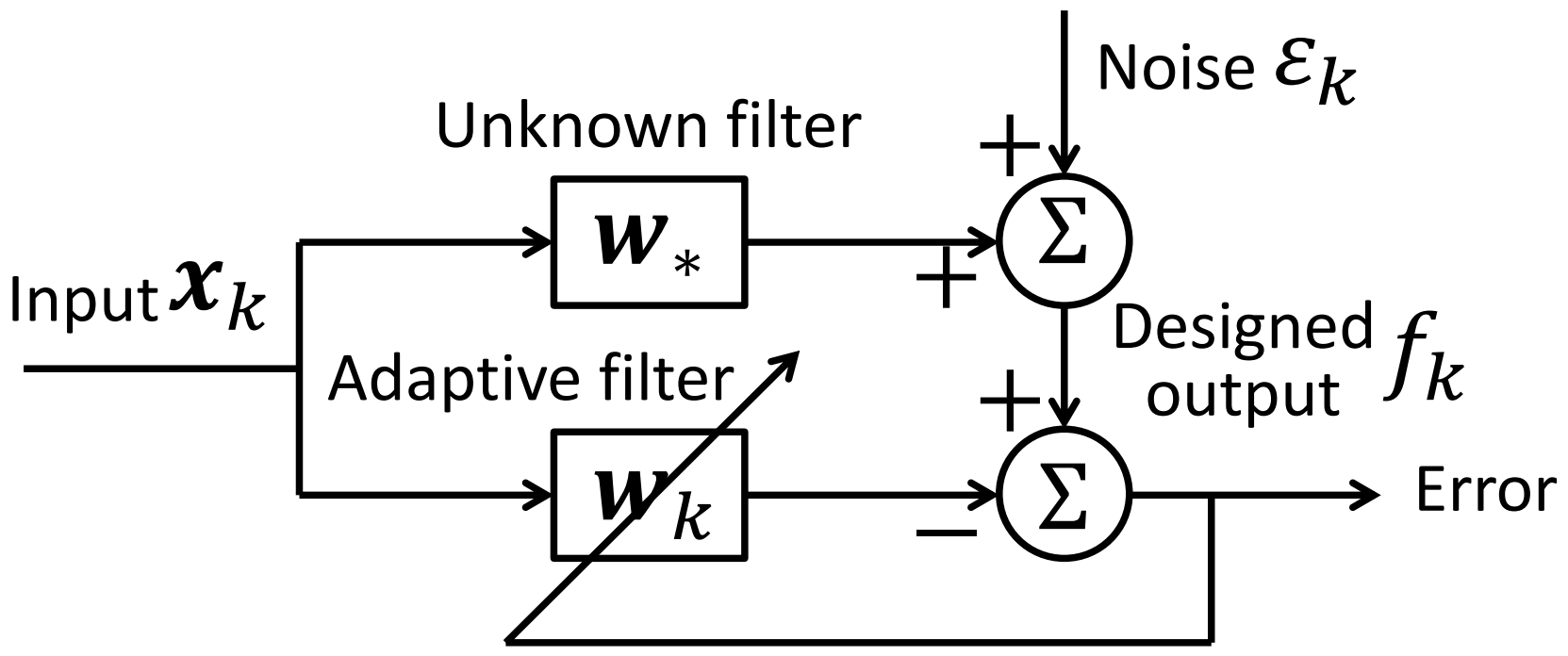

Figure 1. Adaptive filter.

A popular algorithms for adjusting filter weights is the least-mean-square (LMS) [1]:

$$\boldsymbol{w}_{k+1} = \boldsymbol{w}_k + \delta(f_k - \boldsymbol{w}_k^T\boldsymbol{x}_k)\boldsymbol{x}_k, \quad (2)$$

where $\delta$ is the update step size. The popularity of LMS comes from the simplicity of implementation and robustness of performance [1].

The performance of the LMS algorithm can be improved if additional information is available about the statistical properties of the true weight vector, $\boldsymbol{w}_*$ ($f_k = \boldsymbol{w}_*^T\boldsymbol{x}_k + \varepsilon_k$, where the observation noise $\varepsilon_k$, is assumed to be independent of $\boldsymbol{x}_k$). In particular, if the true weight vector is sparse, i.e. contains only a few non-zero weights, the LMS algorithm can be modified to reflect such prior knowledge [2-14]. Specifically, a class of LMS algorithms [11-14] incorporates the sparse prior by introducing a sparsity-inducing $l_p$-norm of the weight vector to the loss function (1) and demonstrates better performance than (2).

Here, we show that the $l_p$-norm constrained algorithms [11-14] can be derived in the online convex optimization (OCO) framework [15-17], which was previously used to derive the standard LMS algorithm (2) in order to account for its robustness [18]. Moreover, we use the OCO framework to derive a novel algorithm for predicting signals generated from sparse weights, which we call online linearized Bregman iteration (OLBI). Compared to the existing algorithms [11-14], the derivation of OLBI relies on a systematic treatment of non-differentiable cost function [19-21] using sub-differentials as is done in convex optimization theory [22]. Although invoking sub-differentials leads to the appearance of an additional step in OLBI compared to standard LMS, it remains computationally efficient. We prove analytically that the steady state performance of OLBI exceeds that of LMS for sparse weight vectors and compare the performance of OLBI numerically to that of recent sparse LMS modifications.

The paper is organized as follows. We start with reviewing the derivation the standard LMS from the OCO framework in Section II. In Section III, we derive OLBI from the OCO framework by using the $l_1$-$l_2$ norm as the sparsity-inducing regularizer. In Section IV we compare OLBI and existing sparse LMS algorithms in the OCO framework. In Section V we prove the convergence of OLBI analytically. We demonstrate the performance of OLBI relative to existing algorithms using numerical simulations. Finally, in Section VII, we conclude and discuss possible future improvements of OLBI.

## II. DERIVATION OF STANDARD LMS FROM ONLINE LEARNING

We start by deriving standard LMS following the OCO framework [15-17]. We update $\boldsymbol{w}_{k+1}$ by minimizing the cost function comprised of a cumulative loss, $\sum_{s=1}^{k}\ell_s(f_s,\boldsymbol{w})$, and a regularizer, $\Psi(\boldsymbol{w})$:

$$\boldsymbol{w}_{k+1} = \text{argmin}_{\boldsymbol{w}}\{\delta\sum_{s=1}^{k}\ell_s(f_s,\boldsymbol{w}) + \Psi(\boldsymbol{w})\}, \quad (3)$$

where $\delta > 0$ will turn out to be the learning rate. The first prediction is generated by initializing the weights, $\boldsymbol{w}_1 = \text{argmin}_{\boldsymbol{w}}\{\Psi(\boldsymbol{w})\}$. Therefore, $\Psi(\boldsymbol{w})$ can be thought to represent our knowledge about $\boldsymbol{w}$ prior to the first time step. The convex cost function $\ell_s$ is usually approximated by a linear form of its gradients [17], $g_k(\boldsymbol{w}_k) = \nabla l_k(\boldsymbol{w}_k)$ and (3) becomes

$$\boldsymbol{w}_{k+1} = \text{argmin}_{\boldsymbol{w}}\{\delta\sum_{s=1}^{k}\langle g_s(\boldsymbol{w}_s), \boldsymbol{w}-\boldsymbol{w}_s\rangle + \Psi(\boldsymbol{w})\}. \quad (4)$$

To derive standard LMS, we choose the regularizer as the $l_2$-norm of the filter weights,

$$\Psi(\boldsymbol{w}) = \frac{1}{2}\|\boldsymbol{w}\|_2^2. \quad (5)$$

Furthermore, we choose the loss at each time step, $\ell_s$, as the prediction error squared and, therefore,

$$g_s(\boldsymbol{w}_s) = \nabla l_s(\boldsymbol{w}_s) = -(f_s - \boldsymbol{w}_s^T\boldsymbol{x}_s)\boldsymbol{x}_s. \quad (6)$$

Substituting (5) and (6) into (4), we obtain

$$\boldsymbol{w}_{k+1} = \delta\sum_{s=1}^{k}(f_s - \boldsymbol{w}_s^T\boldsymbol{x}_s)\boldsymbol{x}_s. \quad (7)$$

Similarly, we obtain $\boldsymbol{w}_k = \delta\sum_{s=1}^{k-1}(f_s - \boldsymbol{w}_s^T\boldsymbol{x}_s)\boldsymbol{x}_s$. After substituting this expression in place of the first k-1 terms of the sum in (7), we find the LMS update rule (2). If we assume that the true filter $\boldsymbol{w}_*$ is constant, and that the input signal $\boldsymbol{x}_k$ is wide-sense stationary, then $\boldsymbol{w}_k$ converges to $\boldsymbol{w}_*$, $\lim_{k\to\infty}E[\boldsymbol{w}_k] = \boldsymbol{w}_*$, if and only if $0 < \delta < 1/\lambda_{\max}$ [1]. Here $\lambda_{\max}$ is the greatest eigenvalue of the input covariance matrix $\boldsymbol{C} = E[\boldsymbol{x}_k\boldsymbol{x}_k^T]$. Therefore the standard LMS algorithm is bias free.

Below, we use a similar approach, along with the sparsity prior on the weights, to derive OLBI, then explore its convergence condition analytically and compare its performance with existing algorithms numerically.

## III. DERIVATION OF ONLINE LINEARIZED BREGMAN ITERATION (OLBI)

To derive OLBI, we use a regularizer, $\Psi(\boldsymbol{w}) = \gamma\|\boldsymbol{w}\|_1 + \frac{1}{2}\|\boldsymbol{w}\|_2^2$), which is the $l_1$-$l_2$ norm of the filter weights also known as the elastic net [23]. By substituting this regularizer in Eq. (4) we obtain:

$$\boldsymbol{w}_{k+1} = \operatorname{argmin}_{\boldsymbol{w}}\{\delta \textstyle\sum_{s=1}^{k}\langle g_s(\boldsymbol{w}_s), \boldsymbol{w} - \boldsymbol{w}_s\rangle + \gamma\|\boldsymbol{w}\|_1 + \frac{1}{2}\|\boldsymbol{w}\|_2^2\}. \quad (8)$$

The optimality condition for the weights in (8) is:

$$\boldsymbol{m}_{k+1} \in \partial\left[\gamma\|\boldsymbol{w}_{k+1}\|_1 + \tfrac{1}{2}\|\boldsymbol{w}_{k+1}\|_2^2\right], \quad (9)$$

where $\partial[.]$ designates a sub-differential [22] and $-\boldsymbol{m}_{k+1}$ is the gradient of the first term in (8):

$$\boldsymbol{m}_{k+1} = -\delta \textstyle\sum_{s=1}^{k} g_s(\boldsymbol{w}_s). \quad (10)$$

Similarly, from the condition of optimality for $w_k$

$$\boldsymbol{m}_{k} = -\delta \textstyle\sum_{s=1}^{k-1} g_s(\boldsymbol{w}_s). \quad (11)$$

By combining (10) and (11), we get

$$\boldsymbol{m}_{k+1} = \boldsymbol{m}_k - \delta g_k(\boldsymbol{w}_k). \quad (12)$$

Substituting (10) into (8) and simplifying, we obtain

$$\begin{aligned}\boldsymbol{w}_{k+1} &= \operatorname{argmin}_{\boldsymbol{w}}\left\{-\langle \boldsymbol{m}_{k+1}, \boldsymbol{w} - \boldsymbol{w}_k\rangle + \gamma\|\boldsymbol{w}\|_1 + \tfrac{1}{2}\|\boldsymbol{w}\|_2^2\right\} \\ &= \operatorname{argmin}_{\boldsymbol{w}}\left\{\tfrac{1}{2}\|\boldsymbol{w} - \boldsymbol{m}_{k+1}\|_2^2 + \gamma\|\boldsymbol{w}\|_1\right\}. \end{aligned} \quad (13)$$

Such minimization problem can be solved component-wise using a shrinkage (soft-thresholding) operation [24, 25],

$$\operatorname{shrink}(a,\gamma) = \begin{cases} a-\gamma, & \text{if } a > \gamma \\ 0, & \text{if } -\gamma \le a \le \gamma \, . \\ a+\gamma, & \text{if } a < -\gamma \end{cases} \quad (14)$$

By combining (6), (12) and (13), we arrive at the following:

**Algorithm 1** Online linearized Bregman iteration (OLBI)

**initialize**: $\boldsymbol{m}_1$=0 and $\boldsymbol{w}_1$=0.

**for** k=1,2,3,…**do**

$$\boldsymbol{m}_{k+1} = \boldsymbol{m}_k + \delta(f_k - \boldsymbol{w}_k^T\boldsymbol{x}_k)\boldsymbol{x}_k, \quad (15)$$

$$\boldsymbol{w}_{k+1} = \operatorname{shrink}(\boldsymbol{m}_{k+1}, \gamma), \quad (16)$$

**end for**

The reason for the algorithm name becomes clear if we substitute (10) and (12) into (8) and obtain:

$$\boldsymbol{w}_{k+1} = \operatorname{argmin}_{\boldsymbol{w}}\left\{\delta\langle g_k(\boldsymbol{w}_k), \boldsymbol{w} - \boldsymbol{w}_k\rangle + \gamma\|\boldsymbol{w}\|_1 + \tfrac{1}{2}\|\boldsymbol{w}\|_2^2 - \langle \boldsymbol{m}_k, \boldsymbol{w} - \boldsymbol{w}_k\rangle\right\}, \quad (17)$$

which can be written as:

$$\boldsymbol{w}_{k+1} = \operatorname{argmin}_{\boldsymbol{w}}\left\{\delta\langle g_k(\boldsymbol{w}_k), \boldsymbol{w} - \boldsymbol{w}_k\rangle + D_{\Psi}^{\boldsymbol{m}_k}(\boldsymbol{w}, \boldsymbol{w}_k)\right\}, \quad (18)$$

where

$$D_{\Psi}^{\boldsymbol{m}_k}(\boldsymbol{w}, \boldsymbol{w}_k) = \Psi(\boldsymbol{w}) - \Psi(\boldsymbol{w}_k) - \langle \boldsymbol{m}_k, \boldsymbol{w} - \boldsymbol{w}_k\rangle \quad (19)$$

is the Bregman divergence induced by the convex function $\Psi(\boldsymbol{w})$ at point $\boldsymbol{w}_k$ for sub-gradient $\boldsymbol{m}_k$ [26].

Table I

Standard and sparse LMS algorithms compared in the OCO framework:

$\boldsymbol{w}_{k+1} = \mathrm{argmin}_{\boldsymbol{w}}\{\delta \sum_{s=1}^{k}\langle g_s(\boldsymbol{w}_s), \boldsymbol{w} - \boldsymbol{w}_s\rangle + \Psi(\boldsymbol{w})\}$ ; $g_s(\boldsymbol{w}_s) = \nabla l_s(\boldsymbol{w}_s)$

| | Loss $\ell_s(f_s, \boldsymbol{w}_s)$ | Regularizer $\Psi(\boldsymbol{w})$ | Recursion |
|---|---|---|---|
| LMS | $\frac{1}{2}(f_s - \boldsymbol{w}_s^T\boldsymbol{x}_s)^2$ | $\frac{1}{2}\|\boldsymbol{w}\|_2^2$ | $\boldsymbol{w}_{k+1} = \boldsymbol{w}_k + \delta(f_k - \boldsymbol{w}_k^T\boldsymbol{x}_k)\boldsymbol{x}_k$ |
| OLBI | $\frac{1}{2}(f_s - \boldsymbol{w}_s^T\boldsymbol{x}_s)^2$ | $\gamma\|\boldsymbol{w}\|_1 + \frac{1}{2}\|\boldsymbol{w}\|_2^2$ | $\boldsymbol{m}_{k+1} = \boldsymbol{m}_k + \delta(f_k - \boldsymbol{w}_k^T\boldsymbol{x}_k)\boldsymbol{x}_k$<br>$\boldsymbol{w}_{k+1} = \mathrm{shrink}(\boldsymbol{m}_{k+1}, \gamma)$<br>$\mathrm{shrink}(z, \gamma) = \max(\lvert z\rvert - \gamma, 0)\mathrm{sgn}(z)$ |
| $l_0$-LMS | $\frac{1}{2}(f_s - \boldsymbol{w}_s^T\boldsymbol{x}_s)^2 + \kappa\|\boldsymbol{w}_s\|_0$ | $\frac{1}{2}\|\boldsymbol{w}\|_2^2$ | $\boldsymbol{w}_{k+1} = \boldsymbol{w}_k + \delta(f_k - \boldsymbol{w}_k^T\boldsymbol{x}_k)\boldsymbol{x}_k + \delta\kappa h^{l_0}(\boldsymbol{w}_k)$<br>$h^{l_0}(z) = \min(\lvert z\rvert - 1/\alpha, 0)\cdot[\alpha^2 z - \alpha\mathrm{sgn}(z)]$ |
| ZA-LMS | $\frac{1}{2}(f_s - \boldsymbol{w}_s^T\boldsymbol{x}_s)^2 + \rho\|\boldsymbol{w}_s\|_1$ | $\frac{1}{2}\|\boldsymbol{w}\|_2^2$ | $\boldsymbol{w}_{k+1} = \boldsymbol{w}_k + \delta(f_k - \boldsymbol{w}_k^T\boldsymbol{x}_k)\boldsymbol{x}_k + \delta\rho h^{ZA}(\boldsymbol{w}_k)$<br>$h^{ZA}(z) = -\mathrm{sgn}(z)$ |
| RZA-LMS | $\frac{1}{2}(f_s - \boldsymbol{w}_s^T\boldsymbol{x}_s)^2 +$<br>$\rho\sum_i \log(1 + \lvert w_{s,i}\rvert/\varepsilon)$ | $\frac{1}{2}\|\boldsymbol{w}\|_2^2$ | $\boldsymbol{w}_{k+1} = \boldsymbol{w}_k + \delta(f_k - \boldsymbol{w}_k^T\boldsymbol{x}_k)\boldsymbol{x}_k + \delta\rho h^{RZA}(\boldsymbol{w}_k)$<br>$h^{RZA}(z) = -\mathrm{sgn}(z)/(1 + \varepsilon\lvert z\rvert)$ |

In fact, (18) can be a starting point for the derivation of OLBI, thus justifying the name. Previously in [19-21], a related algorithm called linearized Bregman iteration was proposed for solving deterministic convex optimization problems, such as basis pursuit [27].

## IV. DERIVATION OF OTHER SPARSE LMS ALGORITHMS

In this section, we briefly review recent sparse LMS algorithms [11-14] which are closely related to our algorithm. We demonstrate that all these algorithms can be derived from the OCO framework with slightly different choices of the loss function and the regularization function in (3) (Table I).

### *A. ZA-LMS and RZA-LMS*

To derive ZA-LMS [11, 12] we start from Eq. (4) but modifying the loss function by adding an $l_1$ term:

$$l_s^{\mathrm{ZA}}(f_s, \boldsymbol{w}_s) = \frac{1}{2}(f_s - \boldsymbol{w}_s^T\boldsymbol{x}_s)^2 + \rho\|\boldsymbol{w}_s\|_1, \tag{20}$$

while keeping the same $l_2$-norm regularizer $\Psi(\boldsymbol{w}) = \frac{1}{2}\|\boldsymbol{w}\|_2^2$. Then if we neglect the non-differentiability of the $l_1$-norm at zero, we can write the weight update as:

$$\boldsymbol{w}_{k+1} = \boldsymbol{w}_k - \delta\nabla l_s^{\mathrm{ZA}}(f_s, \boldsymbol{w}_s). \tag{21}$$

Following [11, 12], we set the gradient of the $l_1$-norm at zero to be zero, which is the value of one of its subgradients. Then we obtain the update rule of ZA-LMS

$$\boldsymbol{w}_{k+1} = \boldsymbol{w}_k + \delta(f_k - \boldsymbol{w}_k^T\boldsymbol{x}_k)\boldsymbol{x}_k + \delta\rho h^{\mathrm{ZA}}(\boldsymbol{w}_k), \tag{22}$$

where the component-wise zero-point attraction function is given by

$$h^{\mathrm{ZA}}(z) = -\mathrm{sgn}(z) = \begin{cases} \frac{z}{|z|} & z \neq 0 \\ 0 & z = 0 \end{cases}. \tag{23}$$

To restrict attraction to smaller weight values, in the reweighted ZA-LMS or RZA-LMS algorithm [11, 12], the uniform attractor (23) is replaced by:

$$h^{\mathrm{RZA}}(z) = -\frac{\mathrm{sgn}(z)}{1 + \varepsilon|z|}, \tag{24}$$

which generates weaker attraction for larger weights. Replacing $h^{\mathrm{ZA}}(z)$ in (22) by $h^{\mathrm{RZA}}(z)$ yields the update rule for RZA-LMS, Table 1, which can be derived from the OCO framework, (4) by using the following loss function:

$$l_s^{\mathrm{RZA}}(f_s, \boldsymbol{w}_s) = \frac{1}{2}(f_s - \boldsymbol{w}_s^T\boldsymbol{x}_s)^2 + \rho \sum_i \log(1 + |w_{s,i}|/\varepsilon). \tag{25}$$

### B. $l_0$-LMS

We can derive $l_0$-LMS [13, 14] by using a loss function with an additional $l_0$-norm

$$l_s^{l_0}(f_s, \boldsymbol{w}_s) = \frac{1}{2}(f_s - \boldsymbol{w}_s^T\boldsymbol{x}_s)^2 + \kappa\|\boldsymbol{w}_s\|_0, \tag{26}$$

Similar to the ZA-LMS and RZA-LMS loss functions (20, 25), the $l_0$-LMS loss function (26) is not differentiable at zero. Following [13, 14], we approximate the non-differentiable $l_0$-norm as a sum of exponentials $\|\boldsymbol{w}_s\|_0 \approx \sum_i[1 - \exp(-\alpha|w_{s,i}|)]$ [28], take the first order term of the Taylor expansion and set the gradient at zero to be zero, thus obtaining

$$\boldsymbol{w}_{k+1} = \boldsymbol{w}_k + \delta(f_k - \boldsymbol{w}_k^T\boldsymbol{x}_k)\boldsymbol{x}_k + \delta\kappa\sum_i[\alpha^2 w_{s,i} - \alpha\mathrm{sgn}(w_{s,i})]. \tag{27}$$

To achieve zero-point attraction, the sign of $w_{s,i}[\alpha^2 w_{s,i} - \alpha\mathrm{sgn}(w_{s,i})]$ should be non-positive. Therefore, as in [13, 14], we replace the last term in (27) by the following zero-point attraction function, $h^{l_0}(\boldsymbol{w}_k)$, which shrinks the small coefficients in the interval $[-1/\alpha, 1/\alpha]$ towards zero:

$$h^{l_0}(z) = \begin{cases} \alpha^2 z - \alpha\mathrm{sgn}(z) & |z| \le 1/\alpha \\ 0 & \text{elsewhere} \end{cases}, \tag{28}$$

and arrive at the $l_0$-LMS update rule [13, 14]:

$$\boldsymbol{w}_{k+1} = \boldsymbol{w}_k + \delta(f_k - \boldsymbol{w}_k^T\boldsymbol{x}_k)\boldsymbol{x}_k + \delta\kappa h^{l_0}(\boldsymbol{w}_k). \tag{29}$$

Because we were able to derive OLBI and other sparse LMS algorithms in the common OCO framework, we are now in a position to compare their operation. First, unlike OLBI where sparsity inducing terms are added to the regularizer, in ZA-LMS, RZA-LMS and $l_0$-LMS they are added to the loss function (Table 1). Because of the zero-point attraction function, the steady state solution in algorithms [11-14] is biased, i.e. $\lim_{k\to\infty} E[\boldsymbol{w}_k] \neq \boldsymbol{w}_*$. In contrast, OLBI is derived by adding the sparsity-inducing terms to the regularizer and introducing a second set of weights related by a continuous shrinkage operation (14). This yielded a two-step iteration, which does not attract the filter weights to zeros directly and generates sparse solution without causing bias in the steady state (Theorem 1).

Second, instead of approximating the non-differentiable sparsity inducing functions or heuristically setting the gradient to zero where it does not exist, OLBI handles the non-differentiability properly using sub-gradients [22].

Next, we prove the convergence of OLBI analytically, Section V and demonstrate its advantage over other sparse LMS algorithms numerically in Section VI.

## V. Convergence analysis

In the following derivations, we assume that the true filter weight vector $\boldsymbol{w}_*$ is constant, and that the input signal $\boldsymbol{x}_k$ is wide sense stationary. We also assume that the zero-mean additive noise $\varepsilon_k$ is independent of $\boldsymbol{x}_k$.

### A. Mean performance

**Theorem 1**. If $0 < \delta < \frac{1}{\lambda_{\max}}$, then $\lim_{k\to\infty} E[\boldsymbol{w}_k] = \boldsymbol{w}_*$, where $\lambda_{\max}$ is the maximum eigenvalue of the input covariance matrix $\boldsymbol{C} = E[\boldsymbol{x}_k\boldsymbol{x}_k^T]$.

**Proof.** Because of the shrinkage operation,

$$\boldsymbol{w}_{k+1} - \boldsymbol{w}_k = \boldsymbol{Q}_k(\boldsymbol{m}_{k+1} - \boldsymbol{m}_k), \tag{30}$$

where $\boldsymbol{Q}_k$ is a diagonal matrix $\mathrm{diag}(\eta_{k,1}, \eta_{k,2}, \ldots, \eta_{k,n})$ with elements $\eta_{k,i} \in [0,1]$,

$$\eta_{k,i} \begin{cases} = 1, & w_{k+1,i}w_{k,i} > 0 \\ = 0, & w_{k+1,i} = w_{k,i} = 0. \\ \in (0,1), & \text{otherwise} \end{cases}$$

Combining (15) and (30) yields

$$\boldsymbol{w}_{k+1} - \boldsymbol{w}_k = \delta(f_k - \boldsymbol{w}_k^T\boldsymbol{x}_k)\boldsymbol{Q}_k\boldsymbol{x}_k. \quad (31)$$

Plugging $f_k = \boldsymbol{w}_*^T\boldsymbol{x}_k + \varepsilon_k$ into (31) and defining the filter misalignment vector as $\boldsymbol{\xi}_k = \boldsymbol{w}_k - \boldsymbol{w}_*$, we obtain

$$\boldsymbol{\xi}_{k+1} = (\boldsymbol{I} - \delta\boldsymbol{Q}_k\boldsymbol{x}_k\boldsymbol{x}_k^T)\boldsymbol{\xi}_k + \delta\varepsilon_k\boldsymbol{Q}_k\boldsymbol{x}_k. \quad (32)$$

Because $\varepsilon_k$, $\boldsymbol{x}_k$, $\boldsymbol{w}_k$ are mutually independent and $E[\varepsilon_k]$=0, taking expectations on both sides of (32) results in

$$E[\boldsymbol{\xi}_{k+1}] = (\boldsymbol{I} - \delta E[\boldsymbol{Q}_k]\boldsymbol{C})E[\boldsymbol{\xi}_k], \quad (33)$$

Since $E[\boldsymbol{Q}_k]$ is a diagonal matrix with elements ∈ [0,1], then from [29] the spectral radius of $E[\boldsymbol{Q}_k]\boldsymbol{C}$ is no greater than that of $\boldsymbol{C}$. Therefore $0 < \delta < \frac{1}{\lambda_{\max}}$ guarantees convergence and the steady state is bias free, i.e. $E[\boldsymbol{\xi}_\infty] = 0$.

### B. *Steady state mean-square deviation (MSD)*

**Theorem 2**. If $0 < \delta < \frac{2}{(\|\boldsymbol{w}_*\|_0+2)\sigma_x^2}$, then the steady state mean square deviation $D_\infty^{OLBI} = \lim_{k\to\infty} E[\boldsymbol{\xi}_k^T\boldsymbol{\xi}_k] = \frac{\delta\sigma_\varepsilon^2\|\boldsymbol{w}_*\|_0}{2-\delta\sigma_x^2(\|\boldsymbol{w}_*\|_0+2)}$, where the $l_0$-norm $\|\boldsymbol{w}_*\|_0$ counts the number of non-zero entries in $\boldsymbol{w}_*$.

**Proof.** Denote the instantaneous mean-square deviation (MSD) as

$$D_k = E[\boldsymbol{\xi}_k^T\boldsymbol{\xi}_k], \quad (34)$$

and the covariance matrix of $\boldsymbol{\xi}_k$ as

$$\boldsymbol{R}_k = E[\boldsymbol{\xi}_k\boldsymbol{\xi}_k^T]. \quad (35)$$

Denote the signal power

$$\sigma_x^2 = E(x_{k,i}^2), \quad (36)$$

and the noise power

$$\sigma_\varepsilon^2 = E(\varepsilon_k^2). \quad (37)$$

Denote

$$U_{nz} = \{i | w_{*,i} \neq 0\},$$
$$U_z = \{i | w_{*,i} = 0\}.$$

Substituting (32) into (35) and after some algebra, we obtain

$$\boldsymbol{R}_{k+1} = (\boldsymbol{I} - 2\delta E[\boldsymbol{Q}_k]\boldsymbol{P}_x)\boldsymbol{R}_k + \delta^2 E[\boldsymbol{Q}_k(2\boldsymbol{P}_x\boldsymbol{R}_k\boldsymbol{P}_x + \boldsymbol{P}_x^2\text{tr}\{\boldsymbol{R}_k\} + \sigma_\varepsilon^2\boldsymbol{P}_x)\boldsymbol{Q}_k^T], \quad (38)$$

where $\boldsymbol{P}_x$ is a $n \times n$ diagonal matrix with diagonal elements equal to $\sigma_x^2$. Noticing

$$D_k = \text{tr}\{\boldsymbol{R}_k\}, \quad (39)$$

we calculate the diagonal elements of $\boldsymbol{R}_{k+1}$ from (38),

$$R_{k+1,i} = \left(1 - 2\delta\sigma_x^2 E[\eta_{k,i}] + 2\delta^2\sigma_x^4 E[\eta_{k,i}^2]\right)R_{k,i} + \delta^2 E[\eta_{k,i}^2](\sigma_x^4 D_k + \sigma_x^2\sigma_\varepsilon^2). \quad (40)$$

Assuming the threshold $\gamma$ is large enough that at steady state $w_{k,i} | i \in U_{nz}$ rarely changes sign, and $w_{k,i} | i \in U_z$ is zero for most of the time, then

$$\lim_{k\to\infty} E[\eta_{k,i}] = E[\eta_{k,i}^2] \approx \begin{cases} 1, \text{if } w_{*,i} \neq 0 \\ 0, \text{if } w_{*,i} = 0 \end{cases}. \quad (41)$$

Combining (39)-(41), we obtain

$$D_{k+1} = (1 - 2\delta\sigma_x^2 + (\|\boldsymbol{w}_*\|_0 + 2)\delta^2\sigma_x^4)D_k + \delta^2\sigma_x^2\sigma_\varepsilon^2\|\boldsymbol{w}_*\|_0. \tag{42}$$

From (42), we arrive at the mean square convergence condition:

$$|1 - 2\delta\sigma_x^2 + (\|\boldsymbol{w}_*\|_0 + 2)\delta^2\sigma_x^4| < 1, \tag{43}$$

which can be simplified to

$$0 < \delta < \frac{2}{(\|\boldsymbol{w}_*\|_0 + 2)\sigma_x^2}, \tag{44}$$

and the steady state MSD,

$$D_\infty^{OLBI} = \frac{\delta\sigma_\varepsilon^2\|\boldsymbol{w}_*\|_0}{2 - \delta\sigma_x^2(\|\boldsymbol{w}_*\|_0 + 2)}. \tag{45}$$

Compared to the steady state MSD of standard LMS [1]

$$D_\infty^{LMS} = \frac{\delta\sigma_\varepsilon^2 n}{2 - \delta\sigma_x^2(n + 2)}, \tag{46}$$

in the OLBI result (45), the filter length $n$ is replaced by the number of non-zeros entries. At the limit of small learning rate,

$$\frac{D_\infty^{OLBI}}{D_\infty^{LMS}} = \frac{\|\boldsymbol{w}_*\|_0}{n}. \tag{47}$$

Therefore, for a sparse filter with $\|\boldsymbol{w}_*\|_0/n \ll 1$, OLBI achieves a much smaller steady state MSD. Note that in the limit of large threshold, $\gamma$, the steady state result (45) does not depend on the threshold, Fig. 3, but the convergence speed does, Fig. 5.

### C. *Instantaneous MSD*

Next, we derive the approximate instantaneous MSD based on the convergence results obtained for steady state, with an additional assumption that $\boldsymbol{x}_k$ is i.i.d. with zero-mean. By substituting $f_k = \boldsymbol{w}_*^T\boldsymbol{x}_k + \varepsilon_k$ into (15), we obtain,

$$\boldsymbol{m}_{k+1} = \boldsymbol{m}_k + \delta(\boldsymbol{w}_*^T\boldsymbol{x}_k + \varepsilon_k - \boldsymbol{w}_k^T\boldsymbol{x}_k)\boldsymbol{x}_k, \tag{48}$$

which can be written separately for each component as

$$m_{k+1,i} = m_{k,i} + \delta(\boldsymbol{w}_*^T\boldsymbol{x}_k + \varepsilon_k - \boldsymbol{w}_k^T\boldsymbol{x}_k)x_{k,i}. \tag{49}$$

To decouple each coordinate, we take the expectation on both sides of (49) and obtain

$$E[m_{k+1,i}] = E[m_{k,i}] + \delta(w_{*,i} - w_{k,i})\sigma_x^2. \tag{50}$$

Then we can calculate the approximate threshold crossing time $\{t_i\}$ for each coordinate. Denote the ordered sequence of the absolute values of the non-zeros entries in $\boldsymbol{w}_*$ from large to small as $\boldsymbol{w}_{*nz}$. Then from (50), the time for the corresponding component of $\boldsymbol{m}$ to reach the threshold is approximately given by

$$t_i \approx \frac{\gamma}{\delta w_{*nz,i}\sigma_x^2} \tag{51}$$

At time between $t_i$ and $t_{i+1}$, the number of components of $\boldsymbol{m}$ which surpass the threshold is $i$, and thus the number of non-zero entries in $\boldsymbol{w}$ is also $i$. Therefore, at $t_i < t < t + 1 < t_{i+1}$, the instantaneous MSD can be written as a summation of the deviation from the zero entries in $\boldsymbol{w}$ which is a constant and the deviation from the non-zero entries:

$$D_t = \sum_{j=i+1}^{\|\boldsymbol{w}_{*nz}\|_0} w_{*nz,j}^2 + \sum_{j=1}^{i} R_{t,j}. \quad (52)$$

Using (40), for the non-zero entries in $\boldsymbol{w}$,

$$R_{t+1,j} = (1 - 2\delta\sigma_x^2 + 2\delta^2\sigma_x^4)R_{t,j} + \delta^2(\sigma_x^4 D_t + \sigma_x^2\sigma_\varepsilon^2). \quad (53)$$

Combining (52) and (53), we obtain

$$D_{t+1} = [1 - 2\delta\sigma_x^2 + (i+2)\delta^2\sigma_x^4]D_t + 2\delta\sigma_x^2(1 - \delta\sigma_x^2)\sum_{j=i+1}^{\|\boldsymbol{w}_{*nz}\|_0} w_{*nz,j}^2 + i\delta^2\sigma_x^2\sigma_\varepsilon^2. \quad (54)$$

Define

$$a_i \equiv 1 - 2\delta\sigma_x^2 + (i+2)\delta^2\sigma_x^4, \quad (55)$$

$$b_i \equiv 2\delta\sigma_x^2(1 - \delta\sigma_x^2)\sum_{j=i+1}^{\|\boldsymbol{w}_{*nz}\|_0} w_{*nz,j}^2 + i\delta^2\sigma_x^2\sigma_\varepsilon^2, \quad (56)$$

we obtain the explicit expression for $D_{t_i<t<t_{i+1}}$,

$$D_{t_i<t<t_{i+1}} = a_i^{t-t_i} D_{t_i} + b_i \frac{1 - a_i^{t-t_i}}{1 - a_i}. \quad (57)$$

After $t = t_i$, $i = \|\boldsymbol{w}_*\|_0$, all the components in $\boldsymbol{m}$ corresponding to the non-zero entries in $\boldsymbol{w}_*$ cross the threshold, and (57) converges to the steady state MSD (45).

## VI. NUMERICAL RESULTS

In this section, we perform numerical simulations to confirm the theoretical results and to demonstrate the effectiveness of OLBI. We consider white input signal $\{x_k\}$ drawn from Gaussian distribution with zero mean and unit variance. The measurement noise $\{\varepsilon_k\}$ is also a zero-mean white Gaussian process with variance $\sigma_\varepsilon^2$ adjusted to achieve SNR=20dB or 10dB. The filter length is set to be $n = 1000$, and the filter sparsity $\|\boldsymbol{w}_*\|_0$ varies from 50 to 1000. The non-zero coefficients in $\boldsymbol{w}_*$ are drawn from a Gaussian distribution with zero mean and unit variance and their locations are randomly assigned. In all numerical experiments, unless mentioned otherwise, the learning rate $\delta = 8 \times 10^{-4}$, the threshold $\gamma = 0.5$, and the filter sparsity is set to $100$. The results are shown as averages of 100 independent trials.

First, we consider the steady state MSD of OLBI with respect to the learning rate $\delta$. Figure 2 shows that the simulation results are consistent with our analytical results.

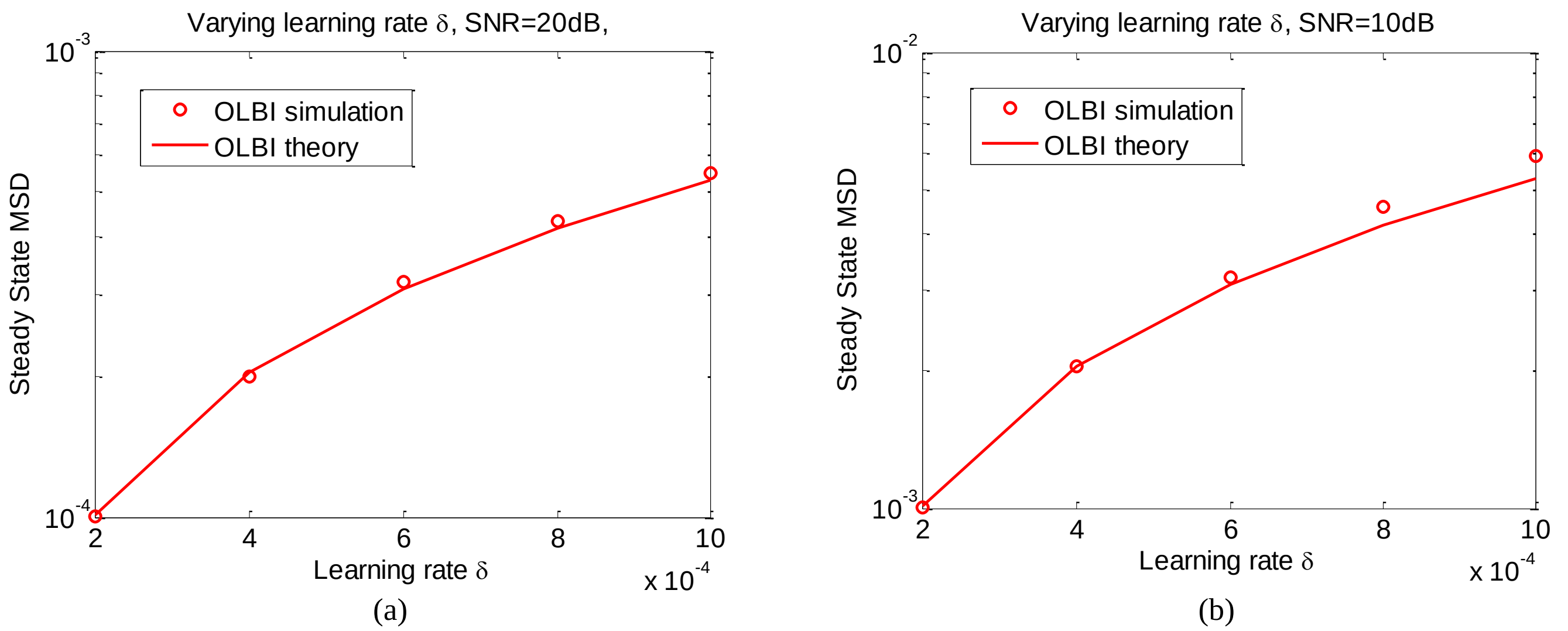

Figure 2. Dependence of steady state MSD on learning rate $\delta$ for SNR=20dB, (a) and SNR=10dB, (b).

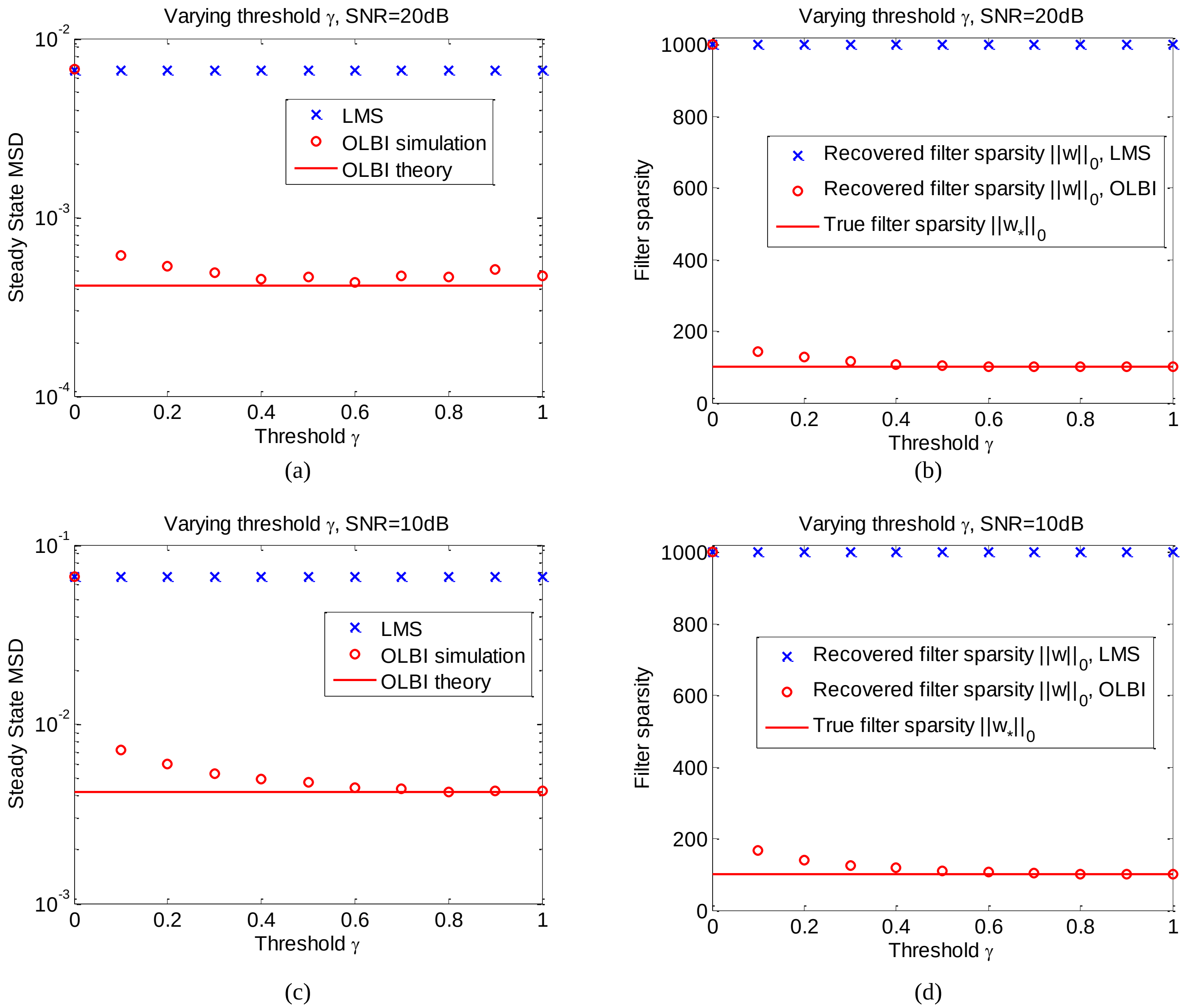

Figure 3. Dependence of steady state MSD and filter sparsity $\|w\|_0$ on threshold $\lambda$ for SNR=20dB, (a) and (b); SNR=10dB, (c) and (d).

Next, we investigate the dependence of steady state MSD on the threshold $\gamma$. The simulation results follow the theoretical predictions well, Fig. 3. When the threshold is large enough, both the steady state MSD (Fig. 3a and c) and the solution sparsity $\|\boldsymbol{w}\|_0$ (Fig. 3b and d) do not depend on $\gamma$. But when $\gamma$ is small, because of noise, zero entries in $\boldsymbol{w}_*$ frequently cross the threshold, yielding solution sparsity and MSD different from true filter sparsity and the theoretical result. Yet, compared to standard LMS, OLBI still achieves smaller MSD. When $\gamma = 0$, OLBI is identical to standard LMS, which is exemplified by the same MSD and solution sparsity $\|\boldsymbol{w}\|_0$, Fig. 3.

In the third experiment, we evaluate the steady state MSD with respect to the system sparsity, Fig. 4. We also included in the comparison other sparse LMS algorithms $l_0$-LMS, ZA-LMS and RZA-LMS with optimized parameters [11-14]. We found the OLBI shows better steady state performance in the tested parameter range thus confirming our theoretical analysis.

The last experiment is designed to investigate the convergence properties of OLBI for different thresholds, $\gamma$, Fig. 5. Smaller thresholds reduce time needed to cross threshold and thus faster convergence rate, but at the expense of denser solutions and larger steady state MSD. When the threshold is large enough, the steady state MSD loses its dependence on $\gamma$. Increasing $\gamma$ only reduces the convergence rate without decreasing the steady state MSD. Comparing numerical results with theoretical predictions with large threshold, $\gamma = 0.5$, we find good general agreement, Fig. 6. The small deviations are due to the imperfect estimation of the timings of threshold crossing events, which are influenced by noise.

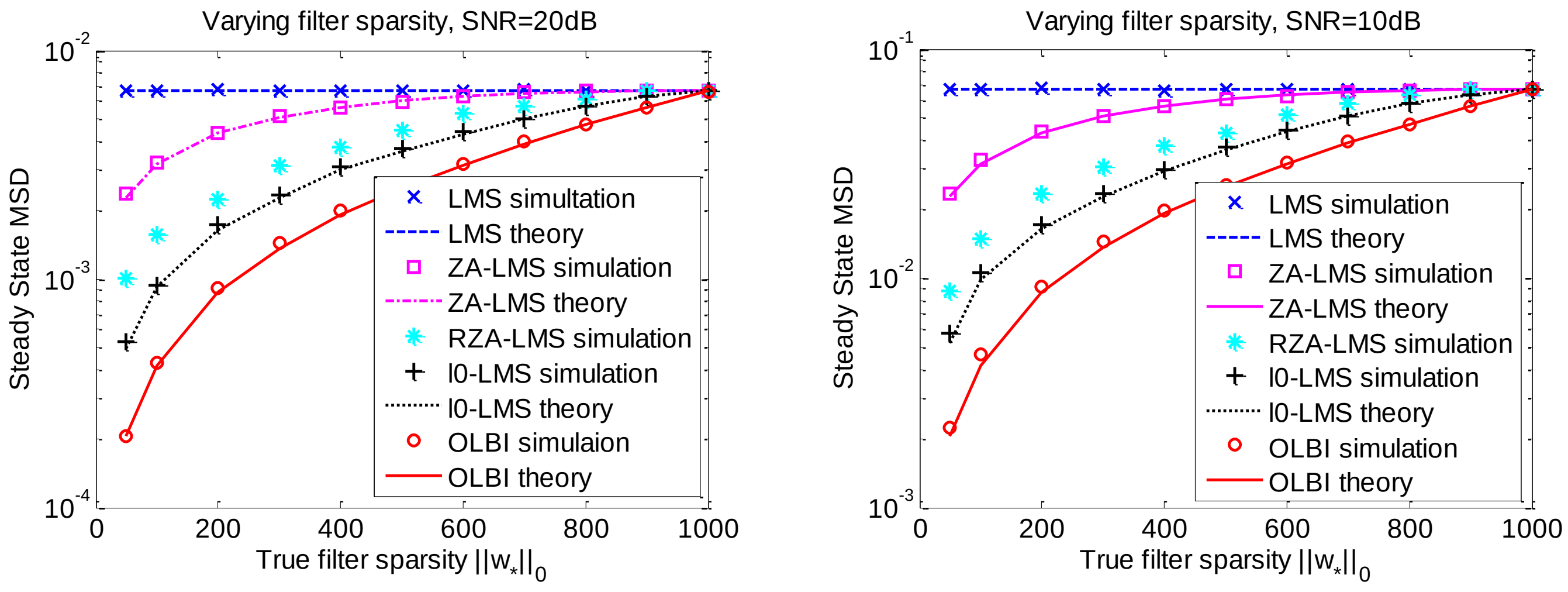

(a) (b)

Figure 4. Dependence of steady state MSD on true filter sparsity for SNR=20dB, (a) and SNR=10dB, (b).

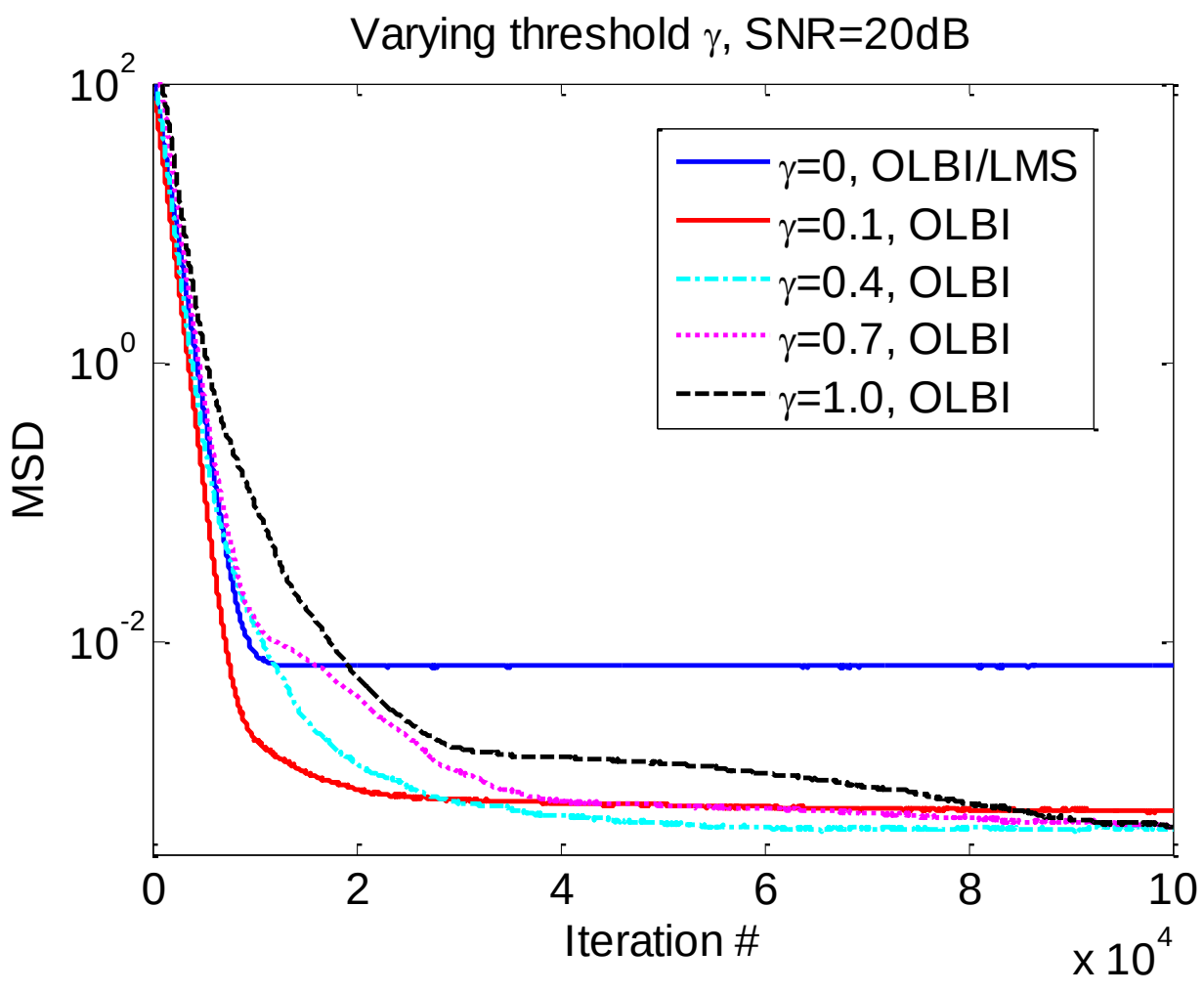

Figure 5. Instantaneous MSD of standard LMS and OLBI with respect to different thresholds.

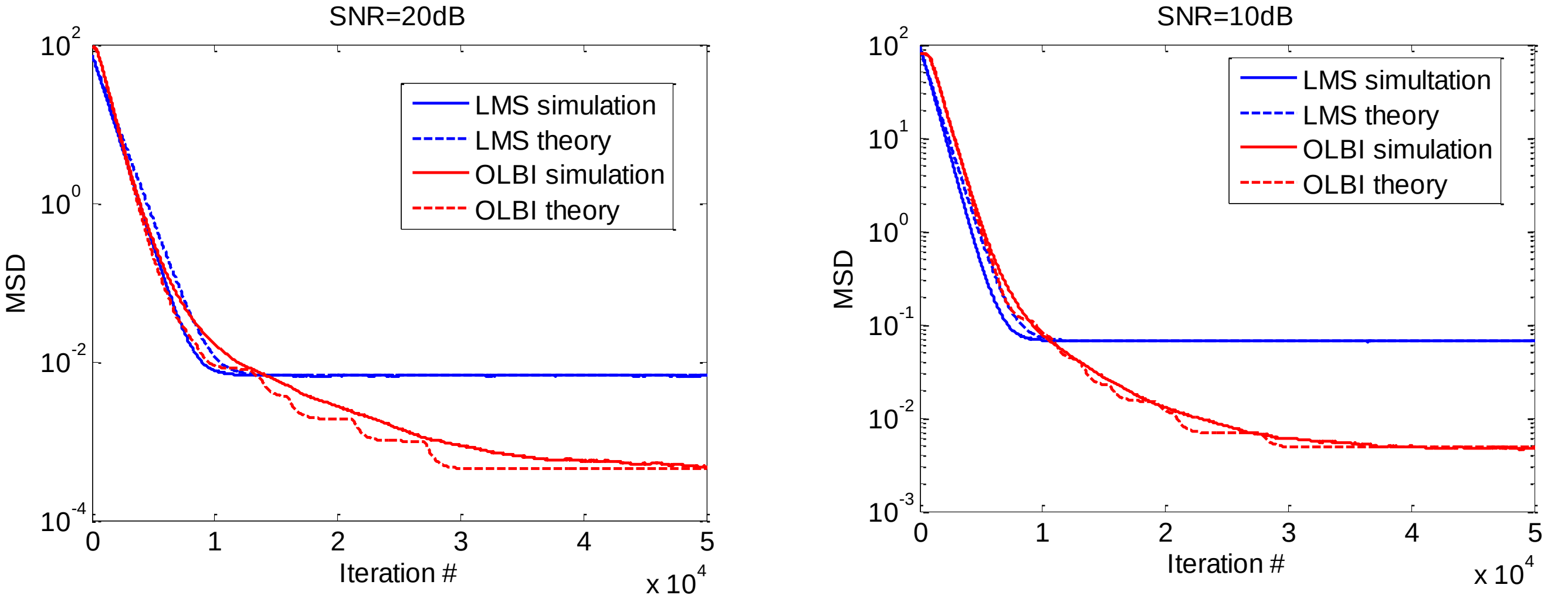

(a) (b)

Figure 6. Instantaneous MSD of standard LMS and OLBI for SNR=20dB, (a) and SNR=10dB, (b).

## VII. Conclusion

We presented a new algorithm called OLBI for solving sparse adaptive signal processing problems. OLBI can be viewed as a sparse version of LMS, which achieves both sparse and unbiased steady state solution. We compared the operation of OLBI with existing sparse LMS algorithms by re-deriving them in the OCO framework. We analyzed its mean square performance. With reasonable approximations, we derived the theoretically results for the steady state and instantaneous MSD. We showed both theoretically and numerically that OLBI is superior to standard LMS for sparse system identification. OLBI also enjoys both low computational complexity and low storage requirement. We may further improve the performance of OLBI by adaptively adjust the learning rate $\delta$, the shrinkage threshold $\gamma$ and incorporate the prior information of structured sparsity for some filters.

## Acknowledgment

We thank Alexander Genkin, Toufiq Parag, Herve Rouault and Karol Gregor for helpful discussions.

## References

[1] B. Widrow and S. D. Stearns, *Adaptive Signal Processing*. New Jersey: Prentice Hall, 1985.
[2] J. Benesty and S. L. Gay, "An Improved PNLMS algorithm," in *Proc. ICASSP*, pp. 1881-1884, 2002.
[3] M. Godavarti and A. O. Hero, III, "Partial update LMS algorithms," IEEE Trans. Signal Process., vol. 53, no. 7, pp. 2382-2399, 2005.
[4] M. Abadi and J. Husoy, "Mean-square performance of the family of adaptive filters with selective partial updates," Signal Processing, vol. 88, no. 8, pp. 2008-2018, 2008.
[5] J. Arenas-Garcia and A. R. Figueiras-Vidal, "Adaptive combination of IPNLMS filters for robust sparse echo cancellation," in *Proc. MLSP*, pp. 221-226, 2008.
[6] H. Deng and R. A. Dyba, "Partial update PNLMS algorithm for network echo cancellation," ICASSP, pp. 1329-1332, Taiwan, 2009.
[7] Y. Murakami, M. Yamagishi, M. Yukawa, and I. Yamada, "A sparse adaptive filtering using time-varying soft-thresholding techniques," ICASSP, pp. 3734-3737, Dallas, TX, 2010.
[8] J. Yang and G. E. Sobelman, "Sparse LMS with segment zero attractors for adaptive estimation of sparse signals," in *APCCAS*, pp. 422-425, 2010.
[9] K. Shi and X. Ma, "Transform domain LMS algorithms for sparse system identification," in *Proc. ICASSP*, pp. 3714-3717, 2010.
[10] Y. Kopsinis, K. Slavakis, S. Theodoridis, and S. McLaughlin, "Online sparse system identification and signal reconstruction using projections onto weighted $l_1$-balls," *IEEE Trans. Signal Proc.*, vol. 59, pp. 936-952, 2011.
[11] Y. Chen, Y. Gu, and A. O. Hero, III, "Regularized least-mean-square algorithms." Available: http://arxiv.org/pdf/1012.5066.
[12] Y. Chen, Y. Gu, and A. O. Hero, III, "Sparse LMS for system identification," in *Proc. ICASSP*, pp. 3125–3128, 2009.
[13] Y. Gu, J. Jin, and S. Mei, "l0 Norm constraint LMS algorithm for sparse system identification," IEEE Signal Process. Lett., vol. 16, no. 9, pp. 774-777, Sep. 2009.
[14] G. Su, J. Jin, Y. Gu, and J. Wang, "Performance Analysis of l0 Norm Constraint Least Mean Square Algorithm." Available: http://arxiv.org/pdf/1203.1535.pdf.
[15] N. Cesa-Bianchi, P. M Long, and M. K. Warmuth, "Worst-case quadratic loss bounds for a generalization of the Widrow-Hoff rule," *IEEE Trans. Neural Networks*, vol. 7, pp. 604–619, May. 1996.
[16] N. Cesa-Bianchi and G. Lugosi, *Prediction, Learning, and Games*. Cambridge: Cambridge University Press, 2006.
[17] A. Rakhlin. (2009). Lecture Notes on Online Learning. Available: http://www-stat.wharton.upenn.edu/~rakhlin/courses/stat991/papers/lecture_notes.pdf.
[18] B. Hassibi, "On the robustness of LMS filters", in Least-Mean-Square Adaptive Filters, S. Haykin and B. Widrow, Eds., John Wiley & Sons, 2003.
[19] J. F. Cai, S. Osher, and Z. Shen, "Convergence of the linearized Bregman iteration for $l_1$-norm minimization," *Mathematics of Computation*, vol. 78, pp. 2127-2136, Oct. 2009.
[20] J. F. Cai, S. Osher, and Z. Shen, "Linearized Bregman oterations for compressed sensing," *Mathematics of Computation*, vol. 78, pp. 1515-1536, Jul. 2009.
[21] W. Yin, S. Osher, D. Goldfarb, and J. Darbon, "Bregman iterative algorithms for $l_1$-minimization with applications to compressed sensing," *Siam Journal on Imaging Sciences*, vol. 1, pp. 143-168, 2008.
[22] S. Boyd and L. Vandenberghe, *Convex Optimization*. Cambridge: Cambridge University Press, 2004.
[23] H. Zou and T. Hastie, "Regularization and variable selection via the elastic net," *Journal of the Royal Statistical* Society *Series B-Statistical Methodology*, vol. 67, pp. 301-320, 2005.
[24] I. Daubechies, M. Defrise, and C. De Mol, "An iterative thresholding algorithm for linear inverse problems with a sparsity constraint," *Communications on Pure and Applied* Mathematics, vol. 57, pp. 1413-1457, 2004.
[25] M. Elad, B. Matalon, J. Shtok, and M. Zibulevsk, "A wide-angle view at iterated shrinkage algorithms," *Proc. SPIE 6701*, 670102, 2007.
[26] L. M. Bregman, "The relaxation method of finding the common point of convex sets and its application to the solution of problems in convex programming," *USSR Computational Mathematics and Mathematical Physics*, vol. 7, pp. 200-217, 1967.
[27] S. S. Chen, D. L. Donoho, and M. A. Saunders, "Atomic decomposition by basis pursuit," *Siam Journal on Scientific Computing*, vol. 20, pp. 33-61, 1998.
[28] J. Weston, A. Elisseeff, and B. Scholkopf, et al, "Use of the zero norm with linear models and kernel methods," the Journal of Machine Learning Research, pp. 1439-1461, 2003.
[29] W. Rudin, *Functional Analysis*. New York: McGraw-Hill, 1991.